\documentclass[journal]{IEEEtran}

\usepackage{graphicx}
\usepackage{color}
\usepackage{placeins}
\usepackage{float}
\usepackage{tabularx,colortbl}
\usepackage{amssymb}
\usepackage{amsthm}
\usepackage{cite}
\usepackage{amsmath}
\usepackage{makecell}
\usepackage{epstopdf}

\usepackage{caption2}

\usepackage{algorithm}
\usepackage{algpseudocode}
\theoremstyle{plain}
\theoremstyle{plain}

\IEEEoverridecommandlockouts

\begin{document}
%----------------------------title&author&thanks----------------------------
\title{Graph Neural Network Meets Multi-Agent Reinforcement Learning: Fundamentals, Applications, and Future Directions}
\author{{Ziheng~Liu, Jiayi~Zhang,~\IEEEmembership{Senior Member,~IEEE}, Enyu~Shi, Zhilong~Liu, Dusit~Niyato,~\IEEEmembership{Fellow,~IEEE}, Bo~Ai,~\IEEEmembership{Fellow,~IEEE}, and Xuemin~(Sherman)~Shen,~\IEEEmembership{Fellow,~IEEE}}
\thanks{Z. Liu, J. Zhang, E. Shi, Z. Liu, and B. Ai are with the School of Electronic and Information Engineering and also with the Frontiers Science Center for Smart High-Speed Railway System, Beijing Jiaotong University; D. Niyato is with Nanyang Technological University; X. Shen is with the Department of Electrical and Computer Engineering, University of Waterloo.}
}
\maketitle

%----------------------------abstract----------------------------

\begin{abstract}
Multi-agent reinforcement learning (MARL) has become a fundamental component of next-generation wireless communication systems. Theoretically,
although MARL has the advantages of low computational complexity and fast convergence rate, there exist several challenges including partial observability, non-stationary, and scalability. In this article, we investigate a novel MARL with graph neural network-aided communication (GNNComm-MARL) to address the aforementioned challenges by making use of graph attention networks to effectively sample neighborhoods and
selectively aggregate messages. Furthermore, we thoroughly study the architecture of GNNComm-MARL and present a systematic design solution.
We then present the typical applications of GNNComm-MARL from two aspects: resource allocation and mobility management. The results obtained unveil that GNNComm-MARL can achieve better performance with lower communication overhead compared to conventional communication schemes. Finally, several important research directions regarding GNNComm-MARL are presented to facilitate further investigation.
\end{abstract}
%----------------------------keywords----------------------------
\IEEEpeerreviewmaketitle
\section{Introduction}
With the rapid development of information technology and the increasing digitization of society, wireless communication, as an important link connecting the world, is undergoing a revolutionary transformation \cite{[14]}. Recently, the sixth-generation (6G) wireless communication system has attracted great attention from both industry and academia, and is expected to lead humanity into a new digital era \cite{[5],[15]}. Compared with previous generations, 6G is expected to achieve faster data transmission, wider network capacity, higher reliability, and lower latency to meet the growing communication needs. Moreover, driven by the novel industrial and technological revolution, 6G will also introduce more intelligent network architectures and innovative communication technologies, bringing more innovative applications and user experiences to society, such as holographic communication, digital twins, and sensor interconnection. However, implementing a 6G network is not a simple task, as it requires in-depth research and innovation in multiple key areas such as wireless communication, network architecture, spectrum management, and device interoperability. Additionally, the design of effective optimization algorithms for 6G faces significant challenges due to its large-scale, low latency, high-density, and integrated multi-functional cross-layer architecture, which results in substantial computational complexity and time delay \cite{[15],[11]}. For instance, the authors in \cite{[16]} investigated a secrecy-energy efficient optimization problem in a satellite-terrestrial integrated network, whose result is of such fundamental importance in this field. However, the computational complexity of the proposed scheme is still relatively high for practical implementation.

Multi-agent Reinforcement learning (MARL) has been recently leveraged as a disruptive technology to solve the challenging combinatorial optimization problems in 6G, as well as support ubiquitous Internet-of-Everything applications including autonomous driving, traffic flow management, drone formation control, sensor networks, and multi-robot collaboration \cite{[1],[12]}. In the MARL framework, all agents interact with the environment to complete tasks, and each agent utilizes RL to learn policies to maximize their cumulative rewards. However, due to the influence of partial observability and non-stationary, agents need to share information with each other, such as states, actions, or experiences, to stably learn and collaborate to complete tasks, conventional MARL algorithms may exhibit poor performance and fail to converge to the optimal policy \cite{[12]}. Therefore, appropriate communication protocols need to be established between agents to communicate and compensate for their partial knowledge of the environment by exchanging their states, actions, and experiences, ensuring that all intelligent agents can collaborate to perform common tasks. Recently, MARL networks with communication (Comm-MARL) have been successfully applied in many scenarios, promoting information sharing among intelligent agents to improve team performance \cite{[5]}. However, most previous work in Comm-MARL failed to capture the complex relationships among agents effectively, resulting in inefficient and low-performing communication.
\begin{table*}[th]
\centering
    \caption{An overview of GNNComm-MARL for wireless communication systems.}
    \vspace{3mm}
    \includegraphics[scale=0.375]{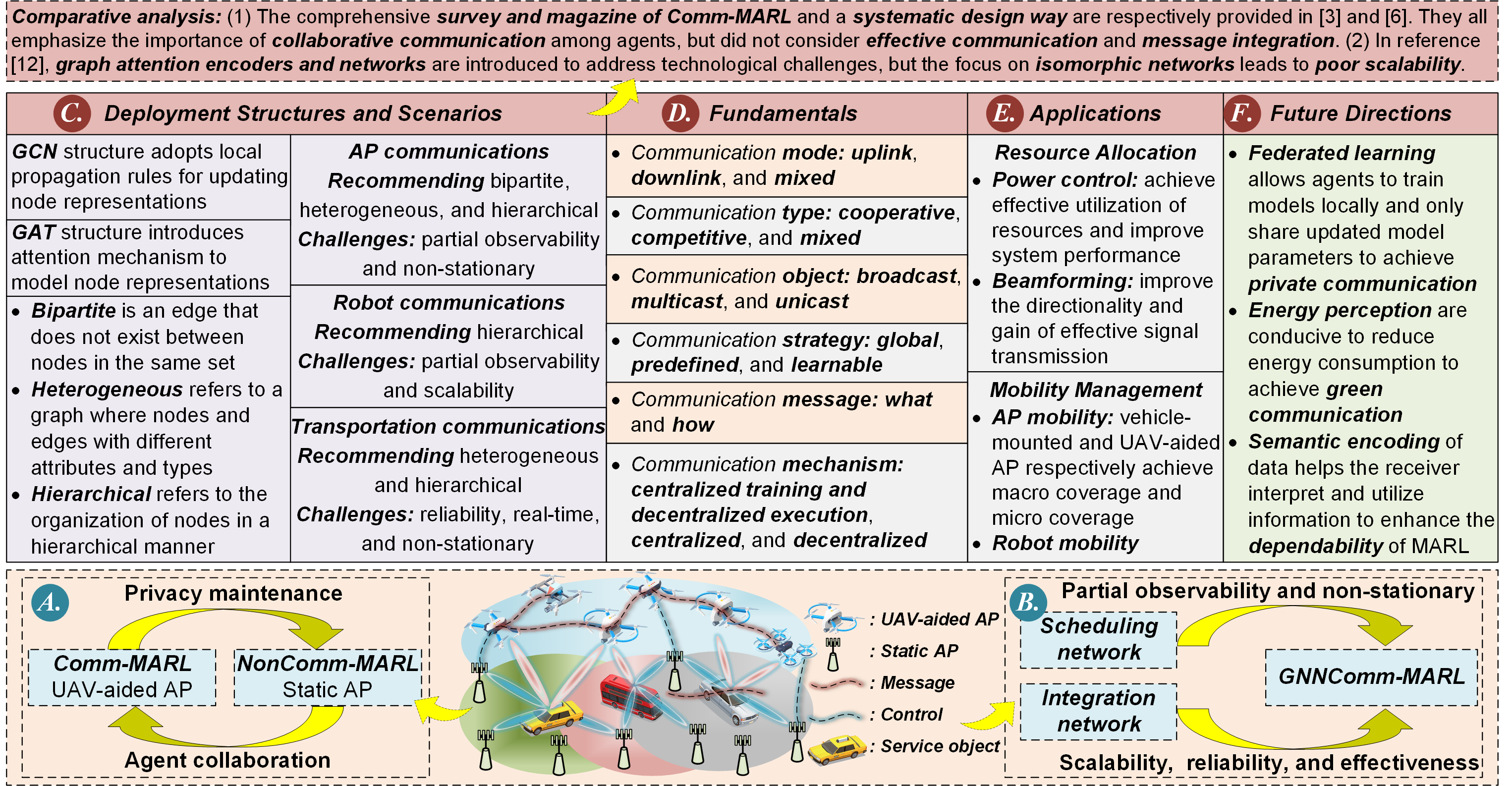}
    \vspace{-3mm}
    \label{fig1:1}
\end{table*}

One promising solution to address these core challenges is to utilize a graph neural network (GNN) for auxiliary communication, modeling the interaction topology between agents as a dynamic directed graph to adapt to the time-varying communication needs among agents and accurately capture their relationships \cite{[6],[8],[9]}. In the MARL network with GNN-aided communication (GNNComm-MARL) \cite{[2],[3],[12]}, the communication protocol can be composed of GNN-aided schedulers and integrators, where the former is responsible for determining when and with whom the agent communicates, while the latter utilizes a graph attention network (GAT) to intelligently combine received messages from neighboring agents, i.e. weighted combination, to achieve high-performance and efficient communication. Specifically, the key idea behind GAT is to assign attention weights to the edges connecting nodes in the graph \cite{[3]}, where the weights are learned through self attention mechanisms rather than predetermined ones. Moreover, the weighted combination mechanism adopted in GAT allows the network to selectively focus on and aggregate node information from neighboring regions, ensuring that each node pays attention to its most relevant neighbors to improve the collaboration ability between agents.

Motivated by the aforementioned works, this paper first provides an overview of GNNComm-MARL for wireless communication systems as shown in Table \uppercase\expandafter{\romannumeral1}. The main contributions are summarized as follows:
\begin{itemize}
\item We analyze the characteristics of conventional MARL and Comm-MARL networks and highlight their architectural differences. Also, we introduce several key challenges faced by these two networks.
\item We discuss three GNN-aided communication structures, including bipartite, heterogeneous, and hierarchical. Then, we propose several deployment scenarios in wireless communication based on features of various GNN-aided communication structures.
\item More importantly, we provide a comprehensive discussion on the framework of GNNComm-MARL from multiple perspectives. Then, numerical results validate the performance of the proposed GNNComm-MARL solutions. Finally, a few novel research directions are highlighted.
\end{itemize}
\vspace{-0.3cm}
\section{Architectures and Challenges of MARL}
In this section, we analyze and compare the two standard architectures of the MARL network, both of which adopt the common multi-agent deep deterministic policy gradient due to its superior applicability, including MARL network with non-communication (NonComm-MARL) and Comm-MARL, as shown in Part A of Table \uppercase\expandafter{\romannumeral1}. Additionally, we discuss the potential challenges that arise from the practical implementation of these two different network architectures, as shown in Part B of Table \uppercase\expandafter{\romannumeral1}.
\vspace{-0.3cm}
\begin{table*}[th]
\centering
    \caption{Comparison of several classical GNN models applied to wireless communication networks in three aspects: attention, aggregation, and combination network, including GCN, GraphSAGE and GAT.}
    \vspace{3mm}
    \includegraphics[scale=0.293]{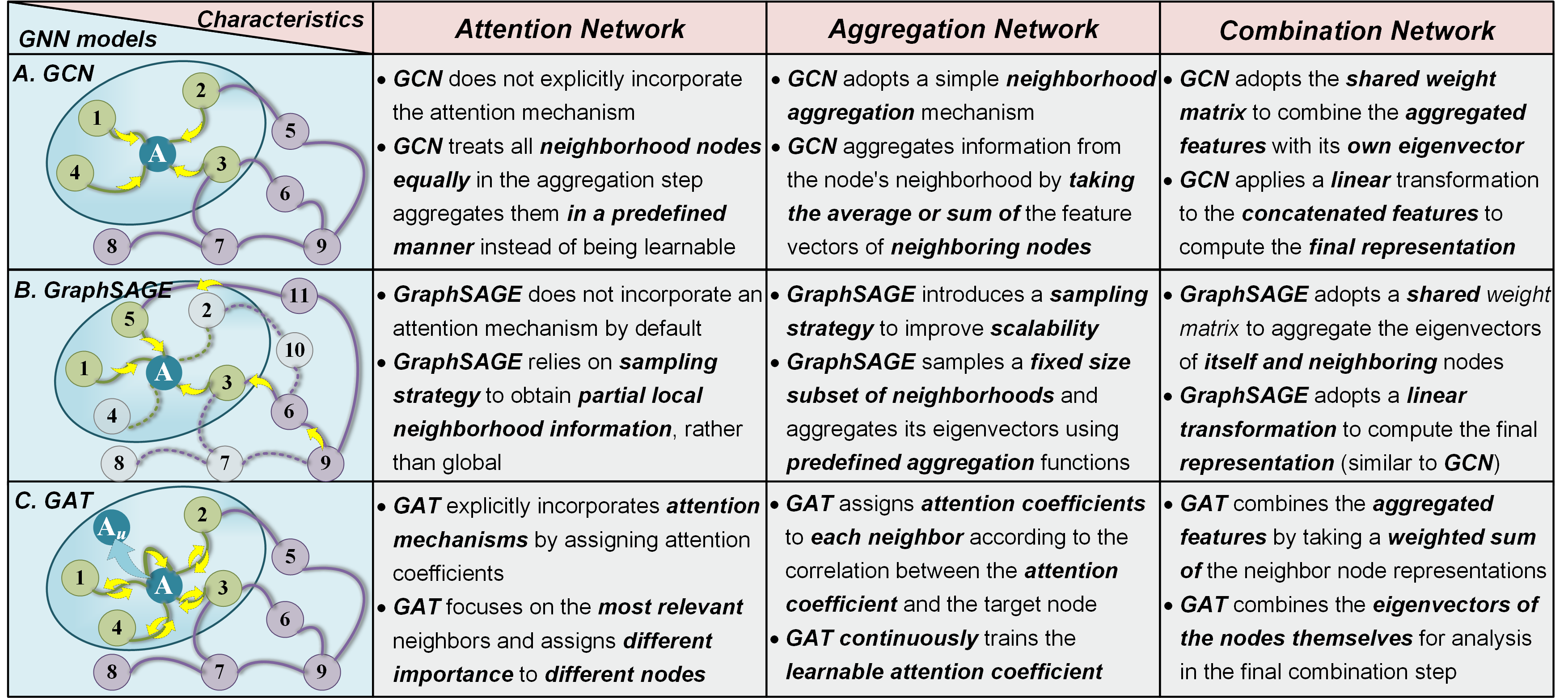}
    \vspace{-3mm}
    \label{fig1:2}
\end{table*}
\subsection{Analysis of NonComm-MARL}
\subsubsection{Network Architecture}
In a multi-agent environment built by NonComm-MARL \cite{[1]}, each agent is an entity (e.g., mobile devices or base stations in wireless networks) that only interacts directly with the environment, rather than interacting with neighboring agents, to effectively avoid message leakage and maintain privacy, i.e., the conversion from Comm-MARL to NonComm-MARL in Part A of Table \uppercase\expandafter{\romannumeral1}. Specifically, although NonComm-MARL models emphasize individual learning and lack information exchange compared to traditional MARL, they still learn and make decisions among multiple agents, making them a specific method (e.g., no communication mechanism) in the field of MARL. Moreover, the observation information of each agent can be regarded as a large number of samples in the continuous state space, and the control of these agents can be approximated as a mapping from a state to an action. In each state, agents take actions given observations of the actual multi-agent environment, and each agent receives an individual reward after interacting with the environment.
%Formally, NonComm-MARL is studied by means of $N$-player Markov decision process (MDP) characterized by a tuple $<\mathcal{S}, \mathcal{A}, \mathcal{P}, \mathcal{R}, \gamma>$, where $\mathcal{S}=(\mathcal{S}^1,\ldots,\mathcal{S}^N)$ is the state space, $\mathcal{A}=(\mathcal{A}^1,\ldots,\mathcal{A}^N)$ is the state space, $\mathcal{R}=(\mathcal{R}^1,\ldots,\mathcal{R}^N)$ denotes the reward functions, $\mathcal{P}:(\mathcal{S},\mathcal{A})\rightarrow\mathcal{S}$ is is the state transition probability, and $\gamma$ is the discounted factor. On each $t$ time slot, agent $i$ receives an observation $o_{t}^{i}=\mathcal{O}(S^i)$ with the observation function $\mathcal{O}$ and maintains an action policy $a_t^i \sim \pi_A^i(\cdot|x_t^i)$ with  its entire trajectory of experience $x_t^i=(o_1^i,a_1^i,\ldots,a_{t-1}^i,o_t^i)$. Then, action policies for each agent are optimized to maximize discounted cumulative joint reward $\mathcal{J}(\pi_A)=\mathbb{E}_{\pi_A}[\sum_{t=1}^{\infty}\gamma^{t-1}r_t^i]$ with ${\pi_A}=\{\pi_A^1,\ldots,\pi_A^N\}$.
\subsubsection{Potential Challenges}
\begin{itemize}
\item \textsl{{Partial Observability:}}
each agent only observes partial environmental information, rather than global, resulting in agents making decisions based on limited observed information, increasing the uncertainty and complexity of decision-making. Therefore, communication should be established among agents to infer unobserved information and consider the implicit intentions of other agents.
\item \textsl{{Non-stationary:}}
the actions of each agent may affect the joint impact of the environment and other agent actions, and their reward depends on their joint decision-making with other agents, increasing the decision-making complexity of the agent. Therefore, agents need to establish appropriate communication protocols to better adapt to changes in non-stationary environments and adjust their strategies by sharing information, i.e., select appropriate actions from the shared observed states.
\end{itemize}
\subsection{Analysis of Comm-MARL}
\subsubsection{Network Architecture}
Comm-MARL is an application of MARL in partially observable Markov games \cite{[5]}, where agents have joint communication channels to address both partially observable and non-stationary environments. By contrast, each agent not only needs to be given partial observations of the actual multi-agent environment in every state but also requires to send messages on shared channels. Each agent learns how to transmit messages based on their experience of interacting with other agents and the environment, and effectively combines the received messages from neighboring agents to take corresponding policies, i.e., the conversion from NonComm-MARL to Comm-MARL in Part A of Table \uppercase\expandafter{\romannumeral1}.
\subsubsection{Potential Challenges}
\begin{itemize}
\item \textsl{{Effective Communication:}}
targetless or uninterrupted communication, in which the agent is not clear when or with whom to communicate, sometimes is not conducive to improving collaboration ability. Therefore, it is necessary to design an effective communication protocol that considers  communication efficiency, interpretability of information, and adaptability to the environment.
\item \textsl{{Reliability:}}
communication between agents may be influenced by information bias and misleading, which may lead to inaccurate, misleading, or biased information being provided, thereby affecting the decisions of other agents. Therefore, the agents need to adopt integrators to reasonably utilize the received information to improve their collaboration capabilities.
\item \textsl{{Scalability:}}
as the number of agents increases, the interaction and collaboration among agents become more complex, requiring more computing resources and time to learn effective policies. Moreover, there may be information bottlenecks and communication costs among agents, which limit the efficiency of information transmission and collaboration. Effectively managing communication and reducing computational complexity are key issues in achieving scalability.
\end{itemize}
\section{Fundamentals of GNNComm-MARL}
In this section, we present a novel GNNComm-MARL paradigm in response to the above challenges and explore its differences from conventional architectures \cite{[2],[3]}. By strategically establishing communication protocols, it becomes possible to effectively combine received messages from neighboring agents to enhance cooperation among agents.
\vspace{-0.3cm}
\subsection{Principles of Graph Neural Network}
Recently, the graph-structured topology of mobile wireless communication systems has enabled them to be successfully utilized to solve a wide range of designed resource allocation problems \cite{[5]}. The emerging GNN can effectively integrate the studied graph structure into the architecture of NNs to model node attributes and relationships between nodes (e.g., mobile devices or base stations), and actively interact between nodes based on potential time-varying relationships to explore hidden features in graph-structured data. Hence, as a specialized NN for graph-structured data, GNN can utilize domain knowledge of various applications (e.g., robot and unmanned aerial vehicle (UAV) communications) to achieve near-optimal system performance through mutual assistance between nodes with good flexibility, generalizability, and scalability \cite{[5],[12]}.
%Traditionally, a graph-structured data can be mathematically represented as a tuple $\mathcal{G} = (\mathcal{V},\mathcal{E})$ of a set of $n_\mathcal{V}$ nodes $\mathcal{V}$ and $n_\mathcal{E}$ edges $\mathcal{E}$. The nodes denote entities of wireless networks (e.g., UEs or APs in cell-free mMIMO communications) and edges represent their relationship (e.g., communication links between UEs and APs). These graph edges can be represented by an adjacency matrix $\mathbf{A} \in \mathbb{R}^{n_\mathcal{V} \times n_\mathcal{V}}$ such that ${A}_{i,j} = 1$ if there exists a communication link between $v_i$ and $v_j$, and 0 otherwise. Moreover, each node $v_i \in \mathcal{V}$ may consist of node features (e.g., beam feature or state information), typically represented through a vector $\mathbf{z}_i$. Similarly, each edge $e_{i,j} \in \mathcal{E}$ between $v_i$ and $v_j$ could also be associated with a set of edge features represented as $\mathbf{e}_{i,j}$ which can, for example, signify channel coefﬁcients.

Typically, GNN models consist of a message-passing scheme with multiple hidden layers, which propagates node features (e.g., channel state information) to their neighboring nodes until a stable equilibrium is reached. In each GNN layer, there are mainly two steps: aggregation and combination, where the former updates the hidden state of each node with its neighbors' information, while the latter aims to combine the obtained aggregated information with the node's own information. Several typical GNN models utilized to improve message passing technology in wireless network topology are shown below, and the comparison can be shown in Table \uppercase\expandafter{\romannumeral2}.
\begin{figure*}[t]
\centering
\includegraphics[scale=0.34]{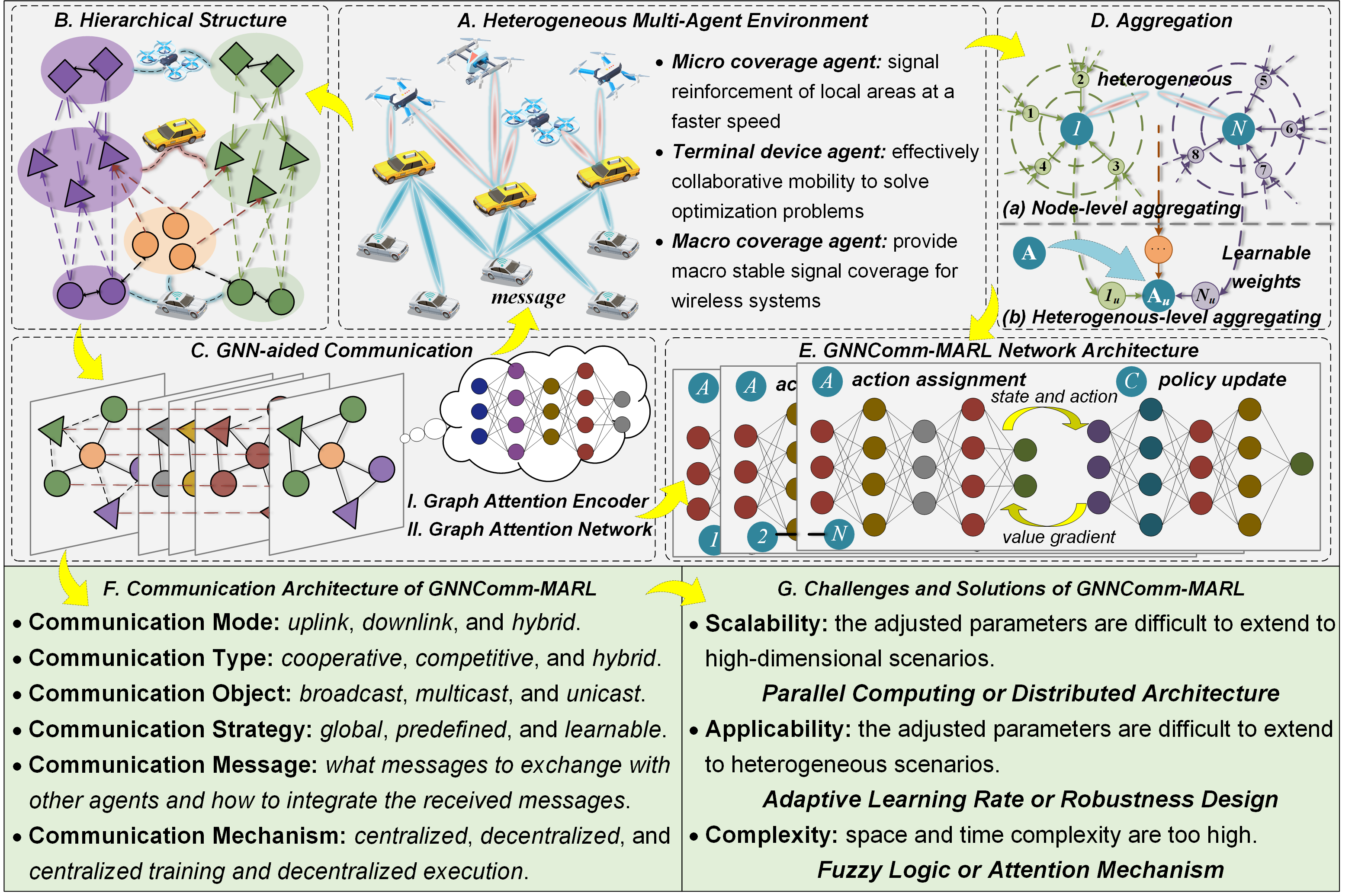}
\caption{The network framework and systematic design scheme of GNNComm-MARL. We improve the original communication protocol by adopting graph attention encoder and GAT, which can adaptively adjust the communication strategy according to the dynamic environment and neighborhood, thus effectively sampling neighbors and aggregating messages.}
\label{fig:3}
\end{figure*}
\subsubsection{Graph Convolutional Network (GCN)}
GCN \cite{[2]} is the first effort that incorporates convolutional operations on GNN. Conventionally, GCN uses convolution operations to aggregate neighbor information of nodes at each layer, which is generated by a linear transformation of node features and application of activation functions. However, the aggregation method of each layer in GCN only considers the first-order neighbor information of nodes, which may not fully capture the high-order relationships among nodes. Moreover, it is important to note that GCN relies on the entire graph (i.e., full adjacency matrix) to learn node representations (i.e., encoding and compression of node features), which is ineffective as the number of neighbors of a node may range from one to thousands or even more.
\subsubsection{Graph Sample and Aggregate (GraphSAGE)}
GraphSAGE \cite{[12]} is an inductive node embedding approach that utilizes node attributes to learn embedding functions, which supports both learning of topological structures and node feature distributions within a confined neighborhood. Compared to GCN, GraphSAGE learns the representation of nodes through sampling and aggregation, making it better able to capture the correlation relationships between nodes and more suitable for large-scale and complex graph datasets. Additionally, GraphSAGE has superior performance in flexibility and scalability through sampling and aggregation policies. Obviously, GraphSAGE has a higher computational complexity $\mathcal{O}(kVE)$ compared to GCN's computational complexity $\mathcal{O}(VE)$, especially when dealing with large-scale graphs (i.e., $k$ is relatively large) that require significant computational resources, where $k$, $V$, and $E$ represent the number of sampled neighbors, nodes, and edges, respectively.
\subsubsection{Graph Attention Network}
GAT \cite{[3]} introduces an attention mechanism that can dynamically learn the importance weights between nodes and neighbors, better capturing the association relationships in the graph structure. Compared to the aforementioned GNN architectures, the contribution of adjacent nodes to the target node is neither predetermined like GCN nor the same as GraphSage. Similarly, the attention mechanism of GAT increases the complexity and computational overhead of the GNN model, which may pose computational resource challenges for large-scale graph datasets.
\subsection{GNN-Aided Deployment Structures and Scenarios}
In this subsection, we propose three typical GNN-aided structures for mobile cell-free massive multiple-input multiple-output (mMIMO) communications, including bipartite, heterogeneous, and hierarchical, which are compared in Part C of Table \uppercase\expandafter{\romannumeral1}. The specific strengths and weaknesses of each structure are outlined below.
\subsubsection{Bipartite}
Most of the current research on cell-free mMIMO systems utilizes the bipartite GNN-aided structure \cite{[7]}. Specifically, we can treat the different communication devices, e.g., users (UEs) and access points (APs), as graph nodes and their interdependent direct links as graph edges, which has the advantage of reduced communication overhead and easy maintenance. However, similar communication devices under the bipartite GNN-aided structure cannot communicate, for example, all APs need to send the received messages to the CPU for centralized processing, which undoubtedly increases system overhead and computational complexity.
\subsubsection{Heterogeneous}
To overcome the shortcomings of the bipartite GNN-aided structure, the heterogeneous GNN-aided structure is proposed by establishing communication links between similar communication devices \cite{[10]}. In the heterogeneous GNN-aided structure, graph nodes can represent different types of communication devices, while graph edges can represent different types of communication relationships, e.g., UE-UE, AP-AP, and AP-UE, breaking the convention limitation of only communicating between AP and UE. This enables more efficient and more flexible deployment of collaborative cell-free mMIMO communications, especially in vehicle-mounted AP or robot communication environments. Obviously, the heterogeneous GNN-aided structure will cause high communication overhead, which also limits the number of communication links and devices. In fact, this heterogeneous GNN-aided structure has poor scalability because it cannot handle communication devices with hierarchical relationships in actual complex communication systems.
\subsubsection{Hierarchical}
To address issues in the above GNN-aided structures, the hierarchical GNN-aided structure establishes a hierarchical structure for subgraph partitioning and node optimization, thereby reducing computational complexity and communication overhead \cite{[4]}. In addition, the hierarchical structure can also support more flexible resource allocation strategies, dynamically adjusting according to the hierarchy to meet the needs of nodes at different levels. However, theoretical research on the hierarchical GNN-aided structure in cell-free mMIMO communications is still immature, and its application in practical scenarios still needs further exploration.

Based on the unique features of each scenario, appropriate GNN-aided communication structures can be selected. For example, when robots, UAV-aided APs, and vehicle-mounted APs collaborate for communication, it is suitable to utilize heterogeneous and hierarchical structures \cite{[4],[10]}. Additionally, for hot spots that require resource management such as beamforming and power allocation, deploying with a bipartite structure is very effective \cite{[7],[9]}.
\begin{figure*}[t]
\centering
\includegraphics[scale=0.235]{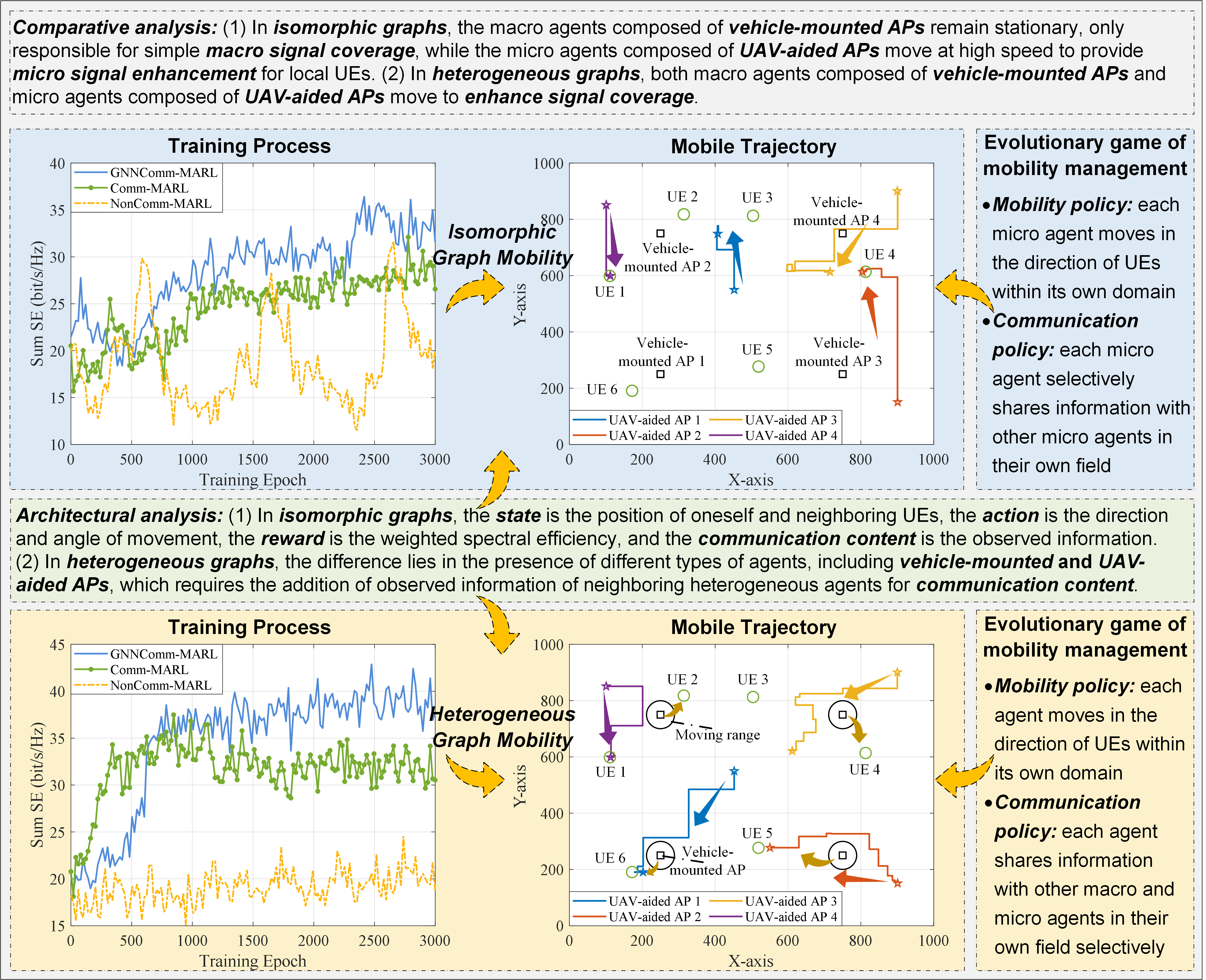}
\caption{The training process and mobile trajectory of GNNComm-MARL in different networks, including isomorphic networks composed of static APs and UAV-aided APs and heterogeneous networks composed of vehicle-mounted APs and UAV-aided APs. We consider a three-slope propagation model adopted in a simulation setup.}
\label{fig:4}
\end{figure*}
\begin{figure*}[t]
\centering
\includegraphics[scale=0.1825]{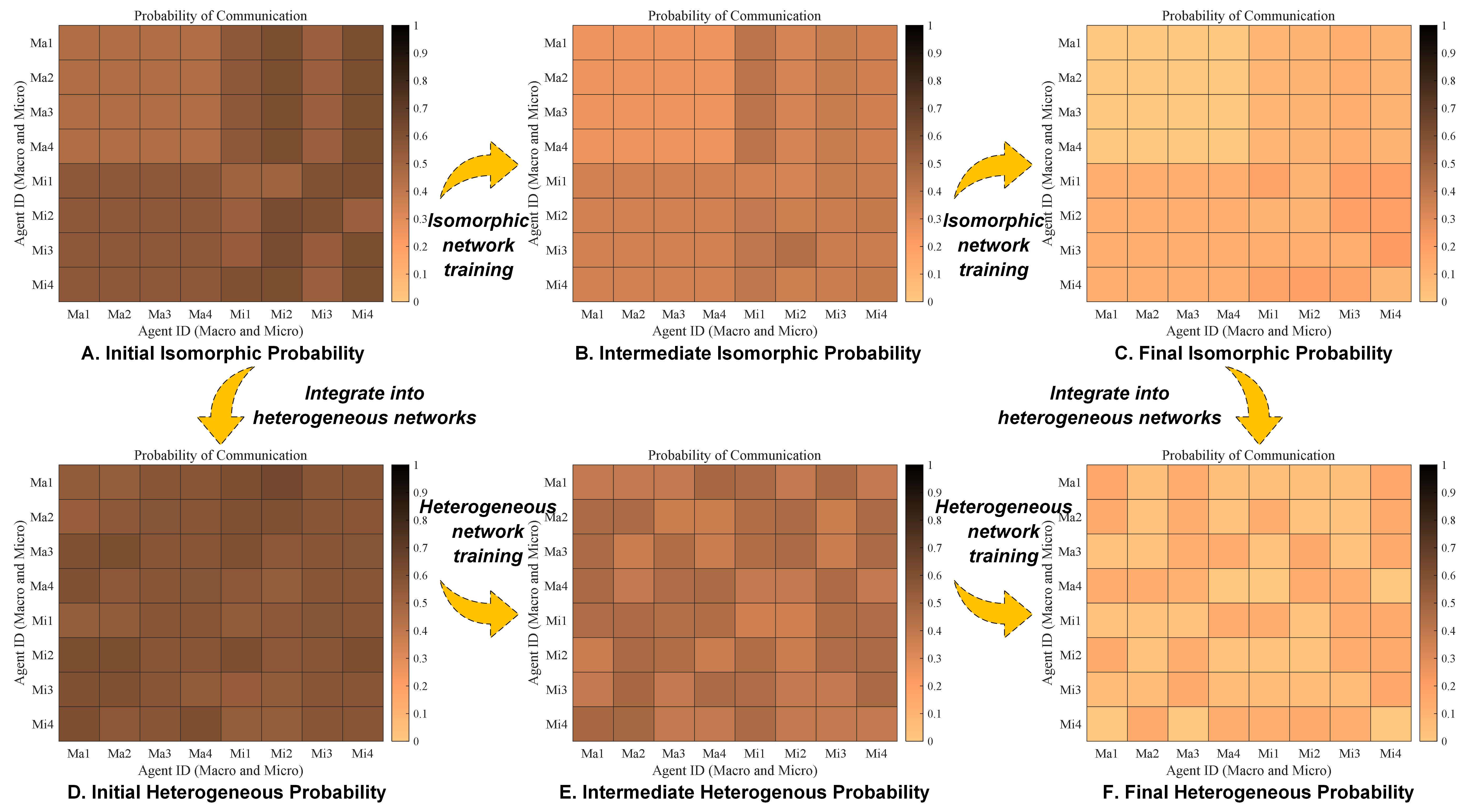}
\caption{Communication probability between agents in two different networks during the training process.}
\label{fig:5}
\end{figure*}
\subsection{Framework of GNNComm-MARL}
Considering the constraints of limited communication resources, low latency frequency, and high capacity, the establishment of communication protocols and the timely exchange of observed information are crucial \cite{[5],[12]}. However, due to the presence of partial observability, conventional communication protocols that transmit fixed messages find it difficult to distinguish between perceptual data related to the target task and local data beneficial to other agents, which poses certain challenges in extracting relevant information for communication between agents. Therefore, it is necessary to adopt advanced collaborative techniques to design adaptable and flexible communication protocols between agents, which has the potential to encode the most relevant and useful information from the original perception data.

Unlike conventional communication protocols that manually encode data for collaboration between agents, GNNComm-MARL adaptively assists agents in learning communication protocols by adopting GNN. In what follows, we investigate a systematic design method of GNNComm-MARL from six aspects, including communication mode and type, to ensure better collaboration among agents while reducing communication overhead, as shown in Fig. 1. Note that algorithms should consider constraints of real-world factors such as limited resources and noise environments, we can alleviate the problem of limited channel capacity by passing short messages including binary. Correspondingly, we can choose a subset of highly correlated agents to adopt shared channels to transmit messages, with the goal of establishing a subset of agents that have strong correlation in terms of tasks and communication requirements, and achieving effective coordination between agents by optimizing the exchange of relevant information. In this network architecture, each agent takes an action and learns a communication message from their observed information and messages transmitted by neighboring agents, where the sending objects (e.g., selected neighborhood agents) of each agent are determined through learning rather than predetermined. Moreover, for messages transmitted by other agents, each agent needs to adaptively determine the correlation between all received messages and tasks to ensure that it can effectively integrate its observed information with received messages.
\begin{figure}[t]
\centering
\includegraphics[scale=0.16]{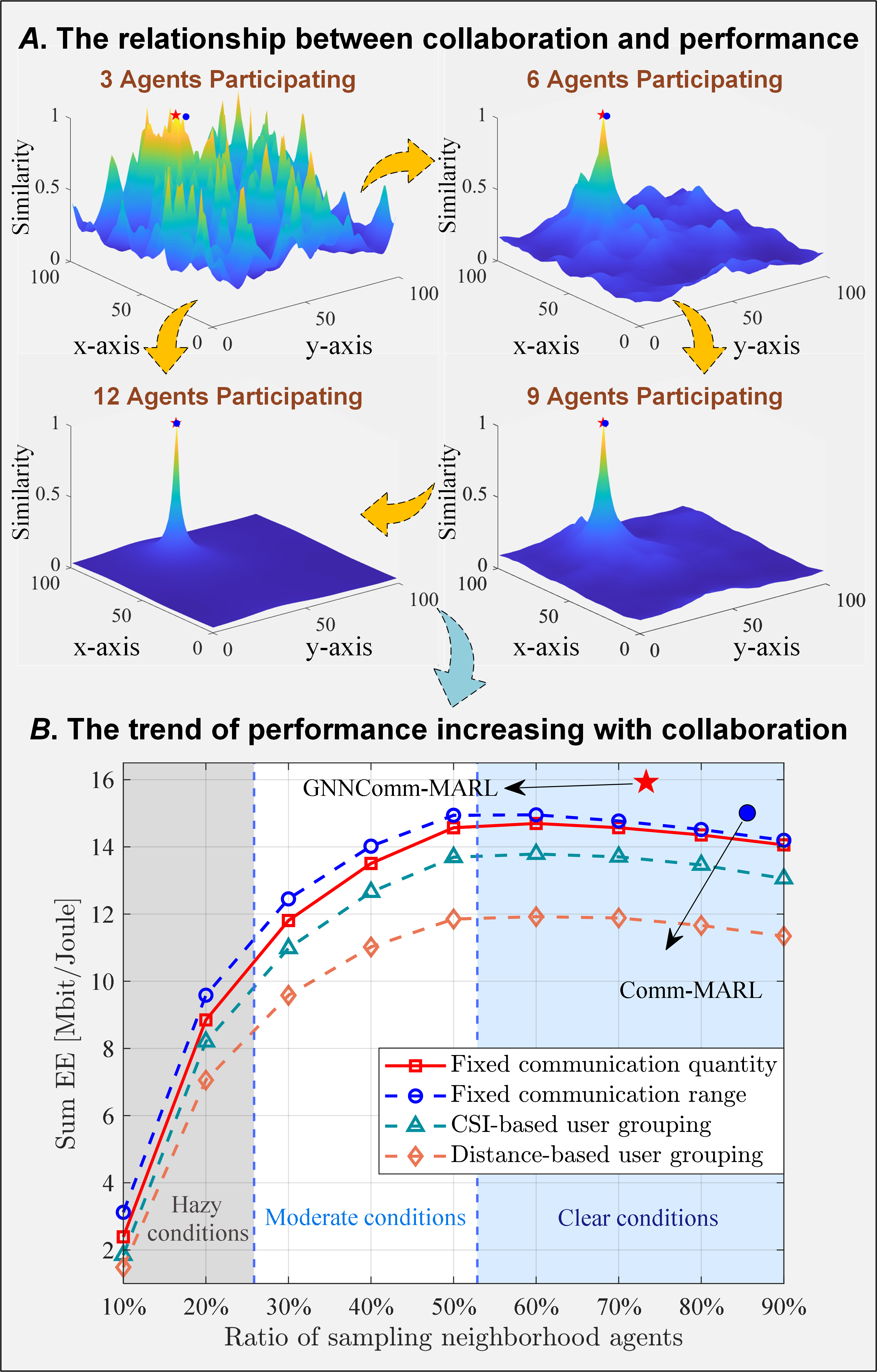}
\caption{The impact of the number of agents participating in collaboration on performance. Figure A illustrates the relationship between collaboration and performance, while Figure B illustrates the trend of performance increasing with collaboration for different communication schemes, where the red star represents GNNComm-MARL and the blue circle represents Comm-MARL.}
\label{fig:6}
\end{figure}
\section{Potential Applications}
Based on the topology of graphs, many emerging applications support 6G that are suitable for adopting the GNNComm-MARL network \cite{[15]}, including cell-free mMIMO and distributed mMIMO, assuming that the same type of agent in each application scenario belongs to the same individual. A typical scenario is to deploy all agents in partially observable environments to collaboratively solve complex tasks. In the following subsections, we take cell-free mMIMO systems as the foundation system model due to its unique capabilities and potential to address several key challenges in the 6G application scenario \cite{[10],[11]}. Moreover, we introduce two potential applications: mobility management and resource allocation, as shown in Part E of Table \uppercase\expandafter{\romannumeral1}, where it is assumed that the mobility state space and resource state space of each agent are the same, but the specific mobility state and resource state are different. Although the assumptions may vary in different application scenarios, they all meet practical requirements. Correspondingly, the relationship between the two can be understood as the change in channel information and interference topology caused by the movement of agents, which indirectly affects the allocation of network resources, making it necessary for the agents to adaptively adjust allocation strategies to improve system reliability and effectiveness.
\subsection{Mobility Management}
Mobility management has long been an essential study topic in developing wireless communication and multi-agent systems, such as AP mobility and robot mobility \cite{[11]}. Due to limitations of partial observability and non-stationarity in practical scenarios, conventional non-cooperative multi-agent networks are no longer applicable. Moreover, considering that communication systems often face challenges such as limited bandwidth resources, the global communication architecture is also not conducive to achieving network scalability. Therefore, when multiple heterogeneous agents are deployed in the field to complete collaboratively common tasks, GNNComm-MARL can serve as a suitable framework to jointly learn the best actions as well as encoded communication messages exchanged between agents.

A clear application is when APs are deployed as mobile base stations to effectively serve UEs and enhance system performance. Each resource-constrained AP needs to optimize its position based on its local observations and message received from other agents. Mobile APs can be divided into vehicle-mounted AP and UAV-aided AP. The former focuses on macro coverage to provide broad wireless network coverage, while the latter focuses on micro coverage to provide larger signal coverage and services for specific areas. In contrast, vehicle-mounted AP has a slower movement speed to provide stable signals for regional UEs, while a UAV-aided AP has a faster movement speed to quickly reinforce signals in weak areas. Note that the above network architecture belongs to a hierarchical graph with multiple heterogeneous networks.

Correspondingly, the evolutionary game of GNNComm-MARL-based mobility management can be described as shown in Fig. 2, we compare the training process of mobile cell-free mMIMO under three schemes. The training process reveals that conventional NonComm-MARL and Comm-MARL have a rapid convergence rate and performance improvement but later fall into local optimums, especially for complex architectures with multi-layer heterogeneous networks. Interestingly, GNNComm-MARL designed based on the GAT has good adaptability in both isomorphic and heterogeneous networks, and can continuously and smoothly converge to the optimal value. The reason is that the introduced GAT network achieves better neighbor sampling and message aggregation, and has better long-term correlation processing capability \cite{[4],[10]}. Additionally, from the mobile trajectory of maximum sum SE, we find that vehicle-mounted APs move slowly within a fixed range to provide stable signals, while UAV-aided APs move faster throughout the entire area to provide signal enhancement to partial special UEs, improving the overall SE performance through communication with each other.
\subsection{Resource Allocation}
Various Comm-MARL-based solutions have recently been designed to address resource allocation issues in cell-free mMIMO systems, e.g. power control and beamforming optimization, where ineffective communication leads to inaccurate received information, further reducing subsequent signal processing capabilities. An effective solution is to adopt a GNN-aided communication network to facilitate collaboration between agents, determine when to communicate, with whom to communicate, and integrate effective received messages, rather than all. For example, the authors in \cite{[3]} developed scheduling networks and integrated networks based on graph attention encoders and GATs, respectively, to enhance collaboration among agents and achieve effective communication, rather than conventional aimless and tedious communication. Unfortunately, their effectiveness is limited to simple graph architectures. For instance, in power control optimization problems, the GNN-aided communication architecture only supports collaboration among APs and cannot achieve simultaneous communication between APs and UEs. A novel GNN-aided communication paradigm known as the heterogeneous graph network is presented to address the demand for wider network scalability with the development of wireless networks. In such a system, collaborative communication among heterogeneous agents composed of AP and UE can be achieved simultaneously, including AP-AP, UE-UE, and AP-UE, to further enhance collaboration capabilities.

Correspondingly, the above network architecture still belongs to a hierarchical graph. Therefore, in Fig. 3, we show the comparison of the communication probability between agents in two different networks during the training process, including homogeneous and heterogeneous networks. For a fair comparison, we normalize the communication probabilities under two different networks.
It is clear that as network training progresses, the communication probability between agents gradually decreases. This reason is that in the initial stage of network training, each agent has limited observation space and requires frequent communication to complete cooperation. However, after multiple training sessions, each agent can better utilize the observation information of the neighborhood, thus avoiding frequent communication and effectively reducing communication overhead.

Additionally, in Fig. 4, we show a performance comparison between conventional communication schemes and GNNComm-MARL by displaying the dynamic relationship between collaboration and performance. Figure A indicates that the system performance improves with the increase of participation, while Figure B indicates that the system overhead also increases with the increase of participation, indicating that it is necessary to choose an appropriate number of agents to participate in collaboration, which is in line with the requirements of real-world scenarios.
We can observe that due to the adaptive collaboration approach adopted by both Comm-MARL and GNNComm-MARL, their EE performance is superior to conventional fixed communication range or quantity schemes. Moreover, it is evident that compared to the conventional Comm-MARL scheme, the proposed GNNComm-MARL scheme can achieve a larger EE performance with lower communication overhead. The reason is that the adopted graph attention encoder and GAT can effectively sample neighborhoods and selectively aggregate messages, respectively, to avoid unnecessary communication overhead and enhance EE performance \cite{[4]}.
\section{Future Directions}
\subsection{Privacy Communications}
In the GNNComm-MARL network, all agents coordinate actions to achieve a common goal by exchanging information, and the communication process may involve the transmission of sensitive information. Information sharing may lead to risks of security and privacy breaches, making protecting communication content a key factor in achieving scalable deployment. Therefore, it is necessary to promote information sharing and cooperation among agents while protecting security and privacy. Federated learning is a possible solution based on distributed learning, which allows agents to train models locally and only share updated model parameters, rather than raw private data. However, federated learning itself also faces some security challenges, such as model information leakage, making the security of applying federated learning in GNNComm-MARL an important open research issue. In contrast, physical layer security can balance security and privacy in wireless communication systems. For example, in cell-free mMIMO systems, beamforming technology is utilized to directional transmit signals to specific users, thereby increasing communication security and reducing the possibility of eavesdropping.
\subsection{Green Communications}
Green communication refers to communication technologies that achieve sustainable development by reducing energy consumption and carbon emissions, but it is worth noting that the relationship between communication perfo2rmance and energy consumption needs to be balanced. In the GNNComm-MARL system, agents may share communication resources and explore how to manage and schedule energy to minimize energy consumption and improve communication efficiency, including resource allocation strategies, design of energy management protocols, as well as conflict resolution and priority scheduling under energy constraints. Moreover, the design of GNNComm-MARL network architecture can consider the characteristics of energy perception to improve energy utilization efficiency.
\subsection{Semantic Communications}
The agents in the GNNComm-MARL network can be enhanced through the implementation of semantic communication, where semantic communication entails the exchange of messages that are not limited to raw source but rather convey semantically understood information. Unlike simple source coding, which focuses on compressing data, semantic communication aims to enable higher-level cooperation and coordination among agents. Furthermore, by utilizing semantic encoding, the transmitted messages carry not only the raw data but also meaningful contextual information and knowledge that enhance the receiver's ability to interpret and utilize the information effectively. This semantic encoding helps to improve the dependability of the GNNComm-MARL system by enabling agents to communicate in a more effective and context-aware manner, facilitating better decision-making and coordination within the network. In addition to semantics, agents can also communicate through other modalities such as vision and sound. By exploring methods of multimodal semantic communication, such as artistic intelligence-generated content, agents can understand semantic information under different perception modes to improve the generalization of design schemes. This may involve techniques such as multimodal data fusion, cross-modal representation learning, and interactive multimodal protocol design.
\section{Conclusions}
In this article, we have proposed the novel GNNComm-MARL paradigm as the main enabling technology for next-generation wireless communication systems and provided three deployment structures and scenarios. A systematic design method of GNNComm-MARL has been presented from six aspects, followed by the respective applications, namely resource allocation and mobility management. Finally, we have discussed three promising
future research directions including privacy communications, green communications, and semantic communications. We hope that this article can provide useful guidance for the design and implementation of GNNComm-MARL in future wireless communication systems.
\bibliographystyle{IEEEtran}
\bibliography{IEEEabrv,Ref}
\end{document}